%% file: main.tex
\documentclass[draft]{agujournal2019}
\usepackage{amsmath,amssymb,bm}
\usepackage[final]{listings} 
\usepackage{array,booktabs,multirow} 
\usepackage[final,hidelinks,implicit=false]{hyperref}
\nolinenumbers
\let\linenumbers\nolinenumbers 
\justifying
\newcommand{\vct}[1]{\boldsymbol{#1}}

\begin{document}
\title{Non-intrusive Differentiation of High-Performance Seismic Solvers}
\authors{Xi Wang\affil{1}, Zhanwen Li\affil{1}, and Xin Liu\affil{1}\thanks{Corresponding author, \href{mailto:liuxine@hku.hk}{liuxine@hku.hk}.}}
\affiliation{1}{Department of Earth and Planetary Sciences, The University of Hong Kong, Pokfulam Road, Hong Kong}

\input{sections/abstract}
\input{sections/introduction}

\input{sections/method}
\input{sections/implementations}

\input{sections/adaptation_workflow}
\input{sections/evaluation}
\input{sections/discussion}

\input{sections/conclusions}
\input{sections/availability}

\bibliography{references}
\end{document}

%% file: sections/abstract.tex
\begin{abstract}
Differentiable frameworks for seismic tomography compute model gradients without hand-derived adjoint sources. 
Some differentiate through wave propagation, which can be costly in time and memory at scale. 
Others use native adjoints with automatically constructed sources, and the efficiency is largely determined by forward and adjoint solves. 
Mature seismic solvers offer efficient, well-tested propagation, but a reusable workflow is still needed to connect them to user-defined misfit functions.
Here we present a non-intrusive workflow and an Agent Skill for adapting existing seismic solvers to differentiable inversion.
We apply automatic differentiation (AD) to the preprocessing and misfit, stopping backpropagation at the synthetic data. 
The solver injects this data derivative at the receivers to form the adjoint source, then computes the model gradient from the native forward and adjoint simulations.
To assess accuracy and computational cost, we compared our hybrid implementation (HybridADFWI) with full AD in ADFWI using the same forward solver.
Across six settings, the gradients agreed to within $1.4\times10^{-6}$ in relative $L_2$ norm.
For one iteration, HybridADFWI was 2.4--4.6 times as fast as ADFWI and used 5--22\% less peak reserved GPU memory.
Guided by the Skill, a coding agent adds missing adjoint routines, verifies gradients, and runs recovery tests for user-selected model parameters.
We applied the Skill to two independent solvers, building DiffSPECFEM2D and DiffAxiSEM3D.
In the P-wave velocity tests, DiffSPECFEM2D recovered the main dipping layers of Marmousi-II, and DiffAxiSEM3D recovered the overall shape of the SEG/EAGE salt body.
Both frameworks support user-defined misfits while preserving native kernels and parallel execution.

\noindent\begingroup\definecolor{abstractlinkblue}{HTML}{376BA2}
GitHub: \href{https://github.com/Wangxi404/differentiate-seismic-solver}{\textcolor{abstractlinkblue}{Agent Skill}},
\href{https://github.com/Wangxi404/DiffSPECFEM-native}{\textcolor{abstractlinkblue}{DiffSPECFEM2D}}, and
\href{https://github.com/Wangxi404/DiffAxiSEM3D-native}{\textcolor{abstractlinkblue}{DiffAxiSEM3D}}.
\endgroup
\end{abstract}

%% file: sections/introduction.tex
\section{Introduction}
\label{sec:introduction}

Gradient-based seismic inversion estimates model parameters by minimizing a misfit function that compares synthetic and observed data.
In full-waveform inversion (FWI), the squared $L_2$ waveform misfit compares amplitudes at corresponding times. When the phase difference between synthetic and observed waveforms is large, minimizing this misfit can lead to cycle skipping.
This limitation has motivated other ways of comparing waveforms \cite{bozdag2011misfit,warner2016adaptive,sun2022mlmisfit}, including traveltime measurements \cite{tromp2005tomography,liu2020differential}, optimal transport \cite{metivier2016transport}, and soft dynamic time warping (soft-DTW) \cite{cuturi2017softdtw,blondel2021divergences,liu2025automatic}.

The adjoint-state method computes model gradients efficiently using an adjoint source that depends on the chosen misfit function \cite{plessix2006adjoint}.
Conventional implementations construct this source from a hand-derived expression for the derivative of the misfit with respect to synthetic data.
The derivation must account for any preprocessing of the synthetic waveforms, such as filtering, windowing, or transformation to another domain.
Changing either the misfit or the preprocessing can therefore require a new derivation, implementation, and numerical check.

Automatic differentiation (AD) allows researchers to obtain misfit derivatives directly from differentiable code \cite{sambridge2007automatic,zhu2021general}.
In full-AD frameworks such as ADFWI, users specify the misfit function in code, and AD differentiates the entire calculation from the model parameters through wave propagation to the misfit \cite{liu2025automatic}.
For long time-domain simulations, however, recording and traversing the computational graph can substantially increase runtime and memory use.
Another established approach uses AD to construct the adjoint source and a native adjoint simulation to compute the model gradient \cite{chen2021hybrid,richardson2026deepwave,louboutin2022abstractions,louboutin2023learned}.
With a native adjoint, each gradient requires both forward and adjoint wave propagation, so inversion speed depends strongly on the underlying solver.

Seismology has accumulated a range of mature, high-performance solvers for forward wave propagation.
These codes combine well-tested numerical methods with parallel implementations optimized over many years.
Examples include the SPECFEM family of spectral-element solvers \cite{komatitsch1998spectral} and AxiSEM3D \cite{leng2019axisem3d,szenicer2020complexity}.
Applications to field data range from imaging the southern California crust \cite{tape2009adjoint} to resolving structure near the core--mantle boundary beneath Hawaii \cite{li2022kilometer}.
Reusing their numerical kernels and parallel implementations avoids much of the validation and optimization needed to build a comparable solver from scratch.

However, the native solvers lack an AD interface for user-defined misfit functions.
Even with an existing adjoint, using a new misfit requires users to derive and implement its adjoint source by hand.
If adjoint routines are missing, they must be implemented from the discrete operations in the forward code. 
The resulting gradients must then be verified.
Therefore, a reusable workflow for making and verifying these extensions across independently developed seismic solvers is needed.

We present a non-intrusive workflow that adapts existing seismic solvers for differentiable inversion while preserving their native numerical kernels and parallel execution (Figure~\ref{fig:solver-bridge}).
Section~\ref{sec:method} shows why AD can stop at the synthetic data. The native solver injects the resulting derivative at the receivers to form the adjoint source and uses its forward and adjoint wavefields to compute the model gradient.
Section~\ref{sec:implementations} describes how our Agent Skill guides a coding agent in reusing or adding solver routines and verifying model gradients.
Section~\ref{sec:evaluation:adfwi} compares our hybrid implementation, HybridADFWI, with full AD in ADFWI to assess gradient accuracy, model recovery, runtime, and GPU memory use.
Section~\ref{sec:evaluation:transfer} tests whether DiffSPECFEM2D and DiffAxiSEM3D, both built using the same Skill, recover known structures in 2D and 3D.
Section~\ref{sec:discussion} discusses how the workflow could extend to other solvers and inverse problems.

\input{sections/solver_bridge_figure}

%% file: sections/solver_bridge_figure.tex
\begin{figure}[tbp]
\centering
\includegraphics[width=\linewidth]{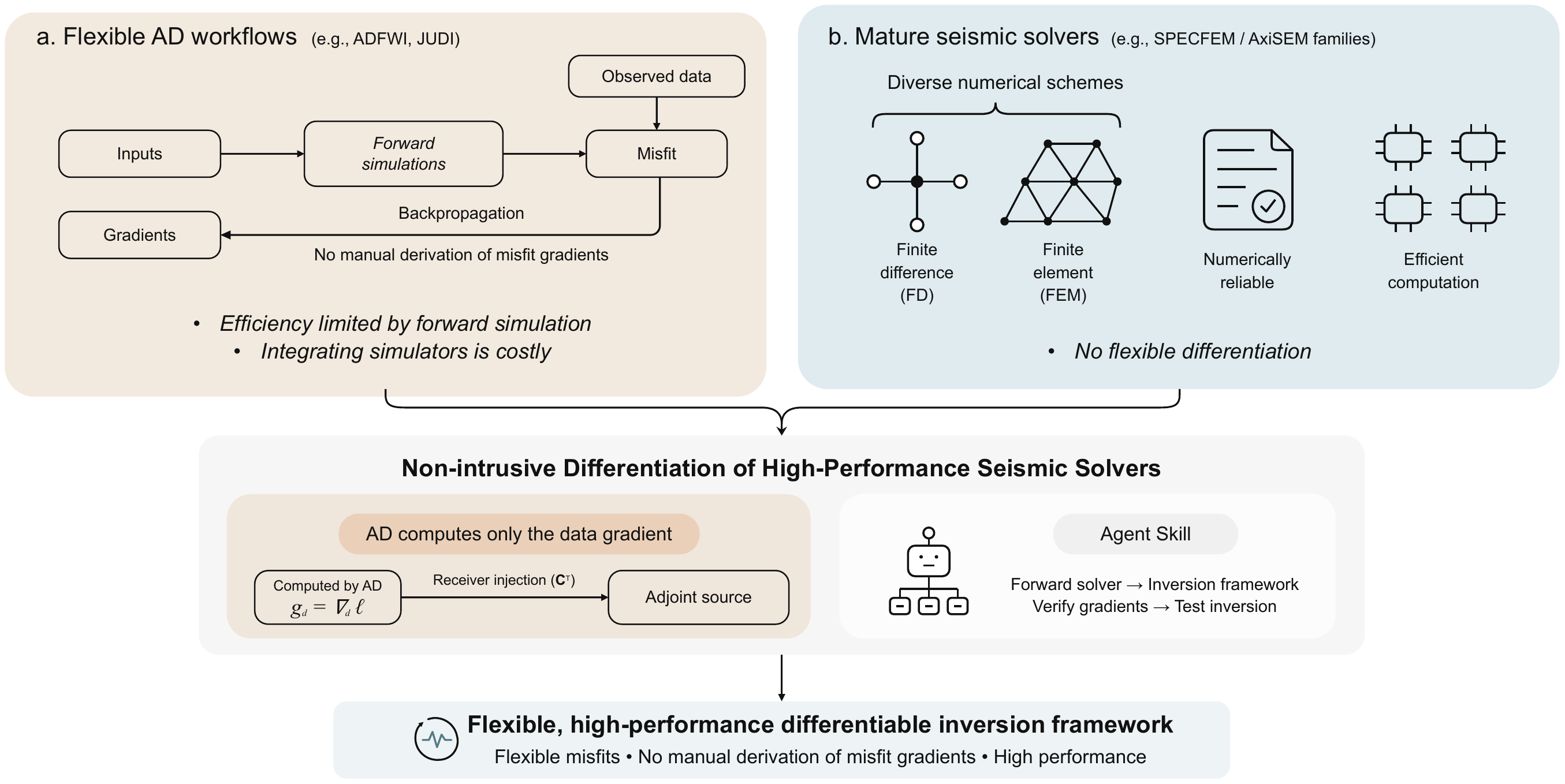}
\caption{Non-intrusive differentiation connects AD workflows (a) with high-performance seismic solvers (b) for inversion. AD computes the derivative of the misfit with respect to synthetic data, $\vct{g}_d=\nabla_{\vct{d}}\ell$. The solver applies the transpose of receiver sampling, $\vct{C}^{\mathsf T}$, to form the adjoint source.}
\label{fig:solver-bridge}
\end{figure}

%% file: sections/method.tex
\section{Non-intrusive differentiation of seismic solvers}
\label{sec:method}

We use AD to differentiate the misfit with respect to synthetic data and the native solver to compute the model gradient (Figure~\ref{fig:gradient-routes}).
The adaptation is non-intrusive because it preserves the solver's numerical kernels and parallel execution while allowing interfaces and adjoint routines to be added.

\input{sections/gradient_routes_figure}

\subsection{Adjoint sources and model gradients}
\label{sec:method:separation}
\label{sec:method:data-derivative}
\label{sec:method:adjoint}

The data misfit depends on the model through the wavefield sampled at the receivers. For model parameters $\vct{m}$ and source $\vct{f}$, the forward wavefield $\vct{u}$ satisfies
\begin{equation}
  A(\vct{m})\vct{u}=\vct{f},
  \label{eq:forward}
\end{equation}
where $A(\vct{m})$ is the discrete wave operator. The receiver matrix $\vct{C}$ samples the wavefield to give the synthetic data
\begin{equation}
  \vct{d}=\vct{C}\vct{u}.
  \label{eq:receiver-sampling}
\end{equation}
The data misfit $\ell$ compares $\vct{d}$ with the observed data $\vct{d}_{\mathrm{obs}}$ and includes any waveform preprocessing. Adding an optional model regularizer $r(\vct{m})$ gives
\begin{equation}
  J(\vct{m})=\ell(\vct{d};\vct{d}_{\mathrm{obs}})+r(\vct{m}).
  \label{eq:objective}
\end{equation}

We seek the model gradient of the data misfit, $\vct{g}_m=\nabla_{\vct{m}}\ell$.
With the source $\vct{f}$ and receiver matrix $\vct{C}$ fixed, the chain rule gives the $j$th component of $\vct{g}_m$ as
\begin{equation}
  g_{m,j}
    =\vct{g}_d^{\mathsf T}\vct{C}
      \frac{\partial\vct{u}}{\partial m_j}.
  \label{eq:gradient-chain-rule}
\end{equation}
Here $\vct{g}_d=\nabla_{\vct{d}}\ell$ has one entry per synthetic data sample, and $\partial\vct{u}/\partial m_j$ describes how the wavefield changes with model parameter $m_j$.
Differentiating equation~\eqref{eq:forward} gives
\begin{equation}
  A(\vct{m})\frac{\partial\vct{u}}{\partial m_j}
    =-\frac{\partial A(\vct{m})}{\partial m_j}\vct{u}.
  \label{eq:forward-sensitivity}
\end{equation}

The adjoint-state method avoids a separate wavefield sensitivity solve for each model parameter \cite{plessix2006adjoint}.
We choose an adjoint wavefield $\vct{\lambda}$ such that $\vct{\lambda}^{\mathsf T}A(\vct{m})=\vct{g}_d^{\mathsf T}\vct{C}$.
Transposing this relation gives the adjoint equation
\begin{equation}
  A(\vct{m})^{\mathsf T}\vct{\lambda}
    =\vct{C}^{\mathsf T}\vct{g}_d
    =\vct{q}_{\mathrm{adj}}.
  \label{eq:adjoint}
\end{equation}
The transpose $\vct{C}^{\mathsf T}$ injects the data derivative at the receiver locations to form the adjoint source $\vct{q}_{\mathrm{adj}}$.

The model gradient can now be evaluated from the forward and adjoint wavefields. Substituting the adjoint relation into the chain rule and using equation~\eqref{eq:forward-sensitivity} gives
\begin{equation}
  g_{m,j}
    =\vct{\lambda}^{\mathsf T} A(\vct{m})\frac{\partial\vct{u}}{\partial m_j}
    =-\vct{\lambda}^{\mathsf T}
      \frac{\partial A(\vct{m})}{\partial m_j}\vct{u}.
  \label{eq:model-gradient}
\end{equation}
At a fixed model and for a given misfit, only the operator derivative $\partial A(\vct{m})/\partial m_j$ changes with $j$ in equation~\eqref{eq:model-gradient}.
For each shot, one forward and one adjoint simulation therefore supply the wavefields for all selected model parameters, such as P-wave velocity, S-wave velocity, or density.
If regularization is included, the full gradient is $\nabla_{\vct{m}}J(\vct{m})=\vct{g}_m+\nabla_{\vct{m}}r(\vct{m})$.

\subsection{AD-based adjoint sources and inversion workflow}
\label{sec:method:implementation}

At a fixed model, the choice of data misfit enters the adjoint equation only through the data derivative $\vct{g}_d$ (equation~\eqref{eq:adjoint}).
Backpropagation can therefore stop at the synthetic data $\vct{d}$ once AD has computed $\vct{g}_d$. The native adjoint and sensitivity routines account for how $\vct{d}$ depends on the model.
The conventional adjoint method uses an analytically derived source (Figure~\ref{fig:gradient-routes}a). In ADFWI, full AD continues through wave propagation to the model parameters (Figure~\ref{fig:gradient-routes}b).

Our workflow connects AD to the native forward and adjoint simulations at the receivers (Figure~\ref{fig:gradient-routes}c). The forward solver computes $\vct{u}$ and returns the synthetic data $\vct{d}$.
PyTorch treats $\vct{d}$ as the input to the differentiable calculation and backpropagates through the misfit and waveform preprocessing to obtain $\vct{g}_d$.
The solver then applies $\vct{C}^{\mathsf T}$ to form $\vct{q}_{\mathrm{adj}}$ and solves equation~\eqref{eq:adjoint} for the adjoint wavefield $\vct{\lambda}$.
Its sensitivity routines use $\vct{\lambda}$ and the saved or reconstructed forward wavefield to evaluate equation~\eqref{eq:model-gradient}. In a time-domain solver, this calculation sums contributions over time for each model parameter.

We compute all shot gradients at the same model and take their mean. We then add any regularization gradient and update the model once.
Algorithm~\ref{alg:inversion-update} shows one complete inversion update.
\begingroup
\renewcommand{\lstlistingname}{Algorithm}
\begin{lstlisting}[language=Python,caption={Full-shot inversion update},captionpos=t,label={alg:inversion-update}]
g_sum = torch.zeros_like(m)
for shot in all_shots:
    with torch.no_grad():
        d, history = native_forward(m, shot)
    d = torch.as_tensor(d).detach().requires_grad_(True)
    loss = data_misfit(d, observed[shot])  # one shot, mean over traces
    g_d, = torch.autograd.grad(loss, d)
    q_adj = receiver_transpose(shot, g_d)
    g_sum += native_adjoint_gradient(m, shot, history, q_adj)
g_m = g_sum / len(all_shots)
g_total = g_m + regularization_gradient(m)  # zero term if unused
update_model_once(m, g_total)
\end{lstlisting}
\endgroup

%% file: sections/gradient_routes_figure.tex
\begin{figure}[tbp]
\centering
\includegraphics[width=\linewidth]{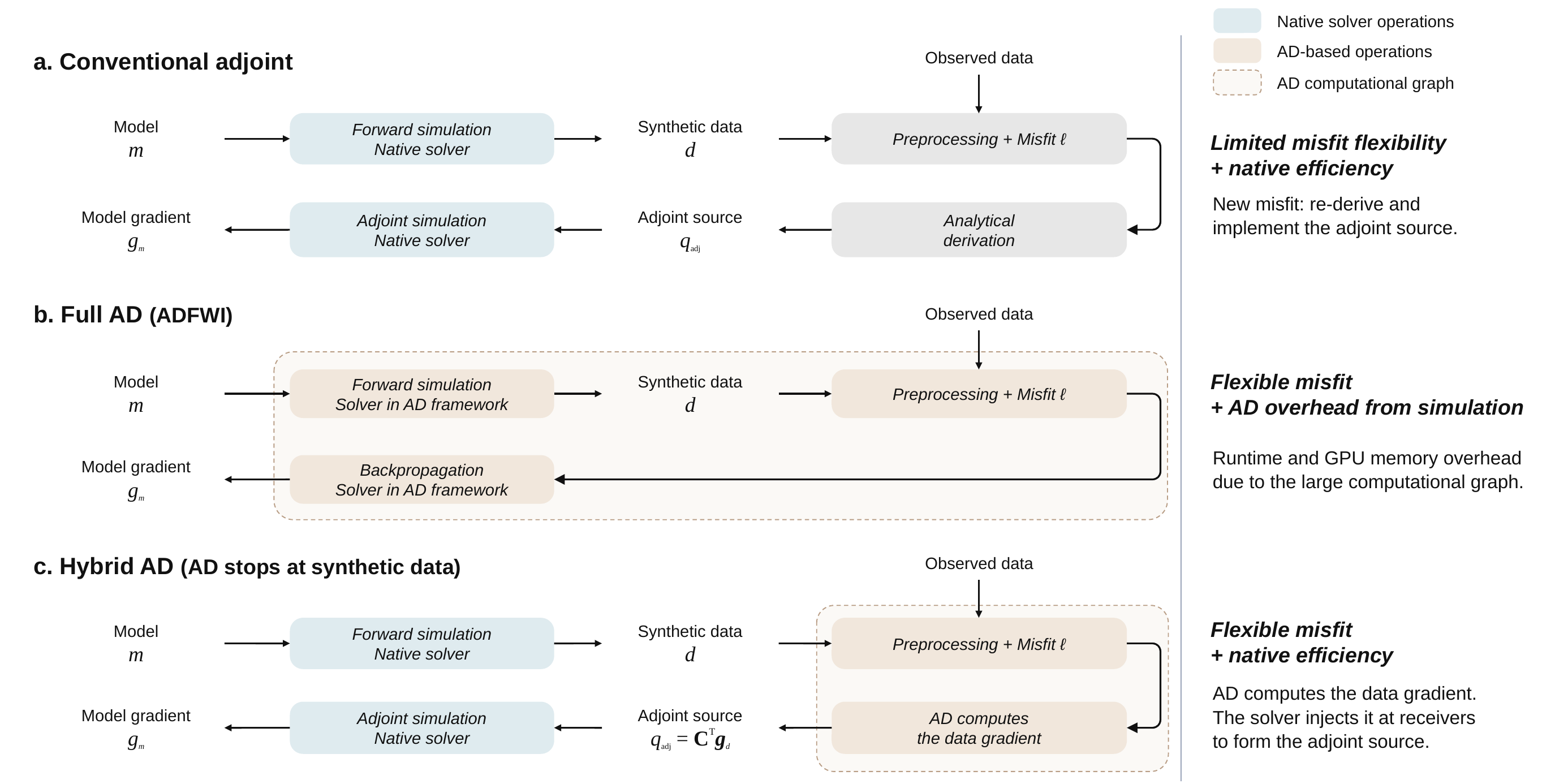}
\caption{The model gradient $\vct{g}_m=\nabla_{\vct{m}}\ell$ can be computed using (a) the conventional adjoint, (b) full AD as in ADFWI \cite{liu2025automatic}, or (c) AD through the misfit with native wave propagation. In (c), AD supplies the data gradient $\vct{g}_d=\nabla_{\vct{d}}\ell$, and the solver injects it at the receivers to form the adjoint source $\vct{q}_{\mathrm{adj}}=\vct{C}^{\mathsf T}\vct{g}_d$.}
\label{fig:gradient-routes}
\end{figure}

%% file: sections/implementations.tex
\section{Adapting seismic solvers with the Agent Skill}
\label{sec:implementations}

The Agent Skill guides a coding agent through building and testing a differentiable inversion interface for an existing forward solver (Figure~\ref{fig:skill-workflow}).
The agent preserves the solver's native numerical kernels and parallel execution as it adds the interface. It then verifies the gradients for the selected parameters and tests model recovery.

\input{sections/skill_workflow_figure}

\subsection{Building and verifying the gradient interface}
\label{sec:adaptation}

To compute gradients for the selected model parameters, the agent identifies which adjoint and sensitivity routines it can reuse and which it must implement.
It inspects how the forward code maps these parameters to material coefficients, advances the wavefield, and samples it at receivers.
The agent reuses an adjoint routine if it implements the transpose of the corresponding discrete forward operation. It reuses a sensitivity kernel if the kernel gives the gradient for the chosen parameter directly or can be converted to that gradient.
The agent implements any missing adjoint or sensitivity calculation from the forward code and identifies the wavefields to save or reconstruct.

The two solvers tested in section~\ref{sec:evaluation:transfer} illustrate how the available native routines determine what the agent must add.
In SPECFEM2D, the agent reused the native adjoint. With $V_{\mathrm S}$ and density held fixed in this test, it converted the bulk-modulus kernel to the required $V_{\mathrm P}$ gradient.
The AxiSEM3D code used here required an adjoint time step, which the agent implemented by reusing the stiffness operator from the forward Newmark scheme.
In both cases, the solver injects the data derivative computed by AD at the receivers to form the adjoint source. The interface returns the resulting model gradient to the optimizer.

Verification starts at the receivers, where the solver returns synthetic data and injects the adjoint source.
The agent first compares the synthetic traces from the adapted solver with those saved from the original forward example under identical settings.
It perturbs the traces to check whether the AD derivative predicts the change in misfit, first for a squared $L_2$ misfit and then for the intended misfit with its preprocessing.
An inner-product test checks that source injection at the receivers is the transpose of wavefield sampling there.

The agent then tests whether the complete model gradient predicts how the misfit changes when the model is perturbed.
For each parameter family being inverted, it applies equal and opposite perturbations along a chosen direction to obtain a centered finite difference of the misfit. It compares this estimate of the directional derivative with the gradient projected onto the same direction.
The Taylor test compares the actual change in misfit with the gradient's linear prediction along the same direction. For a correct gradient, halving the perturbation should reduce the prediction error by about a factor of four over a range of step sizes.
Before the first update, the agent repeats the directional check at the starting model using the same mean over shots as in the inversion.

%% file: sections/skill_workflow_figure.tex
\begin{figure}[tbp]
\centering
\makebox[\linewidth][c]{\includegraphics[width=0.92\linewidth]{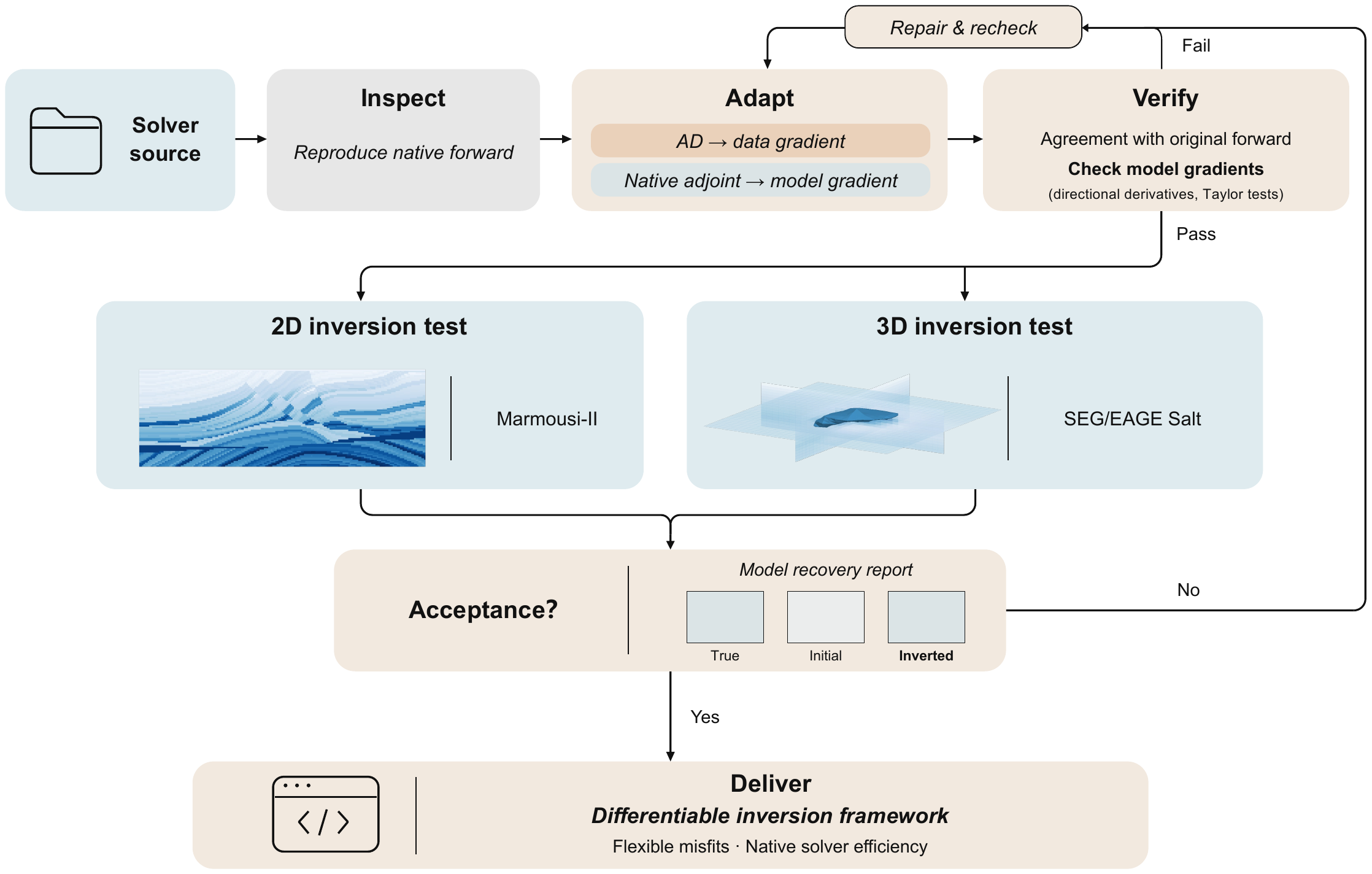}}
\caption{Agent Skill workflow for adapting a forward solver to differentiable inversion. In the Adapt step, the solver injects the data derivative computed by AD at the receivers to form the adjoint source. The default recovery tests target $V_{\mathrm P}$ in Marmousi-II (2D) \cite{martin2006marmousi} and SEG/EAGE Salt (3D) \cite{aminzadeh1997salt}.}
\label{fig:skill-workflow}
\end{figure}

%% file: sections/adaptation_workflow.tex
\subsection{Testing model recovery in 2D and 3D}

Once the gradients have been verified, the agent tests whether successive model updates recover known structure (Figure~\ref{fig:skill-workflow}). The 2D and 3D tests invert for $V_{\mathrm P}$ by default. Users can select one or more parameters supported by the solver, such as $V_{\mathrm S}$ or density, and supply true and starting models for each. For a new selection, the agent reuses or implements the corresponding sensitivity kernels and checks the gradient for each parameter. For joint inversion, it also checks the directional derivative when the selected parameters are perturbed together.

Both recovery tests use synthetic observations generated by the native solver from the true model. In the examples reported here, the inversion starts from a smoothed $V_{\mathrm P}$ model and minimizes the debiased soft-DTW divergence \cite{blondel2021divergences} with $V_{\mathrm S}$ and density fixed.
At each iteration, the agent averages the shot gradients evaluated at the same current model and makes one update.

For a 2D solver, the agent uses Marmousi-II to test recovery of its main dipping layers \cite{martin2006marmousi}.
The Skill samples the University of Houston AGL model on a $200\times68$ grid with 45~m spacing and applies Gaussian smoothing (standard deviation 180~m) to obtain the starting $V_{\mathrm P}$ model. With 40 shots and 200 receivers, the inversion progresses from low frequencies to the full band over at most 300 updates.

For a 3D solver, the agent uses the reduced SEG/EAGE Salt model to test recovery of the salt body's overall shape and velocity contrast \cite{aminzadeh1996models,aminzadeh1997salt}.
The agent maps the $33\times33\times17$ model grid to the solver's native mesh and measures the cost of one gradient evaluation over all shots before inversion. The standard test uses nine shots and 169 receivers over 125 accepted updates.

In both tests, the agent assesses recovery by measuring model error before and after inversion.
The report shows the true, initial, and last accepted models on a common color scale, using the full 2D section or matching horizontal and vertical sections in 3D. It describes which structures the inversion recovered and records why the inversion stopped.

The agent delivers the inversion framework once it passes both gradient verification and the recovery test. If either reveals an implementation error, the agent repairs the extension and repeats the affected checks before delivery. The runnable framework includes the recovery report and the tests and commands needed to reproduce the result.

%% file: sections/evaluation.tex
\section{Verification and applications}
\label{sec:evaluation}

We tested whether restricting AD to adjoint-source construction preserves gradient accuracy and model recovery while reducing runtime and GPU memory use. Full AD in ADFWI provided the reference for this comparison.
We then used the Agent Skill to build DiffSPECFEM2D and DiffAxiSEM3D and tested whether these frameworks based on native solvers could recover known velocity structure in 2D and 3D.

\subsection{Full AD versus AD for the adjoint source}
\label{sec:evaluation:adfwi}

We built HybridADFWI on ADFWI's forward solver so that both methods solved the same wave propagation problem. ADFWI differentiates through wave propagation to obtain the reference model gradient \cite{liu2025automatic}. In HybridADFWI, AD stops at the synthetic data. The solver injects the resulting data derivative at the receivers to form the adjoint source and computes the model gradient with the discrete adjoint of the same time-stepping scheme.
Both implementations used single-precision wave propagation and the same Marmousi-II model \cite{martin2006marmousi}, synthetic observations, and misfit function.

The shared forward problem describes coupled compressional (P) and vertically polarized shear (SV) waves in an isotropic medium. ADFWI and HybridADFWI solve this problem with the same fourth-order finite-difference scheme. The governing plane-strain equations also apply to the SPECFEM2D example in section~\ref{sec:evaluation:transfer} \cite{komatitsch1998spectral}. The in-plane displacement $\vct{u}=(u_x,u_z)^{\mathsf T}$ and stress $\vct{\sigma}$ satisfy
\begin{equation}
  \begin{aligned}
    \rho\,\frac{\partial^2\vct{u}}{\partial t^2}-\nabla\cdot\vct{\sigma}
      &= \vct{f},\\
    \vct{\sigma}
      &= \lambda_{\mathrm L}(\nabla\cdot\vct{u})\vct{I}
       + \mu\bigl[\nabla\vct{u}+(\nabla\vct{u})^{\mathsf T}\bigr].
  \end{aligned}
  \label{eq:psv-displacement}
\end{equation}
Here $\vct{f}$ is the equivalent force density corresponding to the discrete source in equation~\eqref{eq:forward}, $\rho$ is density, and $\vct{I}$ is the in-plane identity tensor. The Lam\'e parameters are $\lambda_{\mathrm L}=\rho(V_{\mathrm P}^2-2V_{\mathrm S}^2)$ and $\mu=\rho V_{\mathrm S}^2$.

We chose soft-DTW divergence \cite{blondel2021divergences} because each synthetic sample can contribute to several time alignments, making the adjoint source difficult to derive by hand. For preprocessed synthetic and observed traces $\vct{p}$ and $\vct{o}$, the divergence is
\begin{equation}
  D_\gamma(\vct{p},\vct{o})
  =S_\gamma(\vct{p},\vct{o})
   -\tfrac12 S_\gamma(\vct{p},\vct{p})
   -\tfrac12 S_\gamma(\vct{o},\vct{o}).
  \label{eq:softdtw-divergence}
\end{equation}
Here $S_\gamma$ is the soft minimum over alignment costs \cite{cuturi2017softdtw}. We used squared amplitude differences with $\gamma=0.1$. For each shot, we divided each trace divergence by its processed sample count and averaged over receivers. Both methods then averaged over all 40 shots, regardless of batch size. AD accounts for preprocessing and differentiates both terms that depend on $\vct{p}$, including both arguments of $S_\gamma(\vct{p},\vct{p})$.

At the smoothed starting model, the complete $V_{\mathrm P}$ gradient from HybridADFWI closely matched the full-AD reference. Across the six settings in Table~\ref{tab:hybridadfwi-benchmark}, the largest relative $L_2$ difference was $1.35\times10^{-6}$. With all 40 shots in one batch and checkpointing enabled, the two implementations also produced identical waveforms.

\begin{figure}[tbp]
\centering
\makebox[\linewidth][c]{\includegraphics[width=1.08\linewidth]{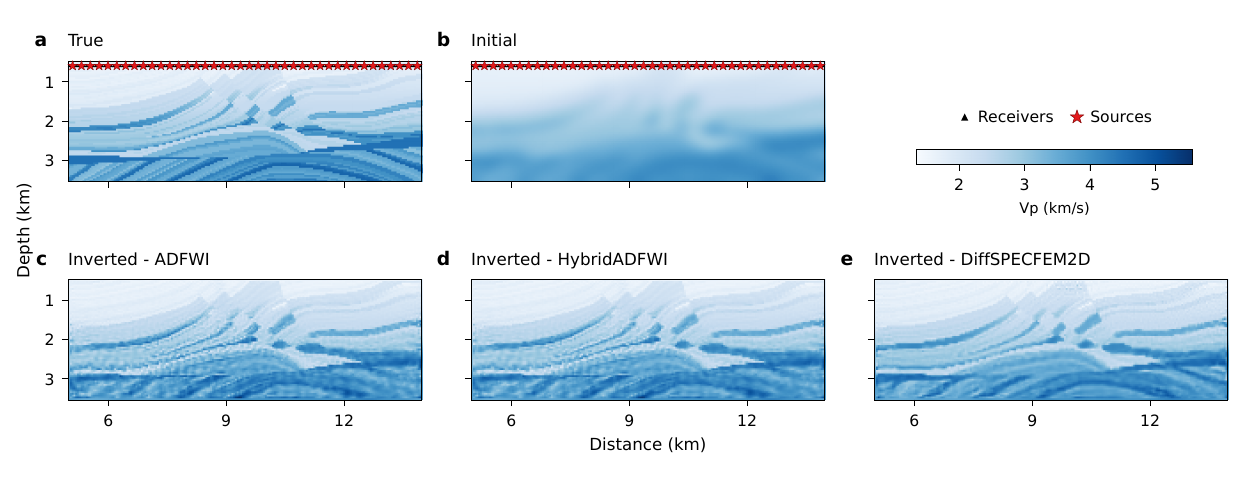}}
\caption{$V_{\mathrm P}$ recovery on Marmousi-II \cite{martin2006marmousi} after 300 updates. From the same starting model (b), full AD in ADFWI (c) and AD only for the adjoint source in HybridADFWI (d) recover the main dipping layers of the true model (a). DiffSPECFEM2D (e), built with the Skill, recovers these layers from SPECFEM2D's own synthetic displacement observations.}
\label{fig:marmousi2-models300}
\end{figure}

This gradient agreement carried over to model recovery. Starting from the same smoothed model and using the same optimizer, both methods recovered the main dipping layers of Marmousi-II after 300 updates (Figure~\ref{fig:marmousi2-models300}a--d).

Restricting AD to adjoint-source construction also reduced the cost of a forward and gradient evaluation in all six configurations. For one evaluation over all 40 shots on the NVIDIA GB10, HybridADFWI was 2.4--4.6 times as fast as ADFWI and used 5--22\% less peak reserved GPU memory across the tested batch sizes and checkpointing settings (Table~\ref{tab:hybridadfwi-benchmark}).

\input{sections/hybridadfwi_checkpoint_comparison}

\begin{figure}[tbp]
\centering
\makebox[\linewidth][c]{\includegraphics[width=0.9\linewidth]{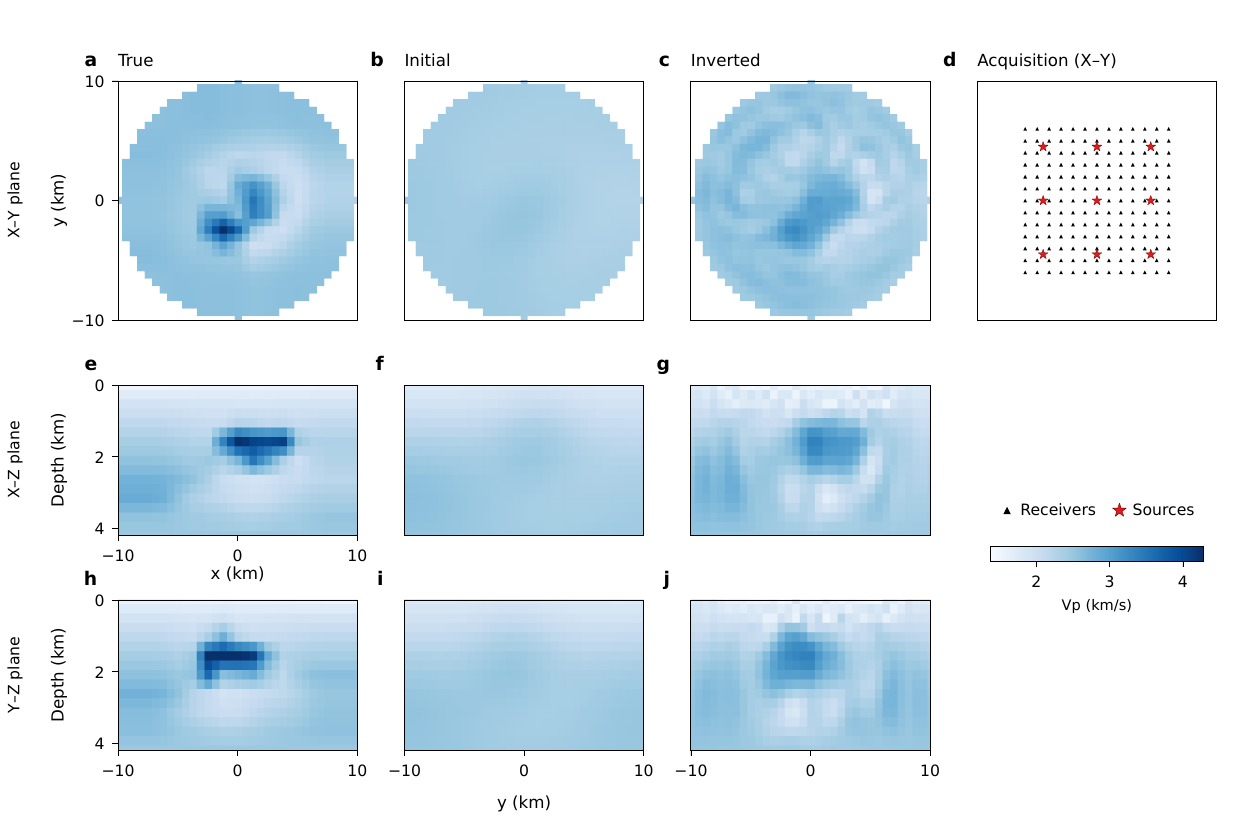}}
\caption{DiffAxiSEM3D recovers the overall shape of the SEG/EAGE salt body \cite{aminzadeh1997salt}. Matching sections show the true (a,e,h), initial (b,f,i), and inverted (c,g,j) models at 2.09~km depth (a--c), $y=0$ (e--g), and $x=0$ (h--j). 
Vertical sections have threefold vertical exaggeration.
Panel (d) shows sources (stars) and receivers (triangles).}
\label{fig:axisem3d-models125}
\end{figure}

\subsection{Building DiffSPECFEM2D and DiffAxiSEM3D with the Agent Skill}
\label{sec:evaluation:transfer}
\label{sec:evaluation:objectives}
The ADFWI comparison tested the method on a single forward solver. We next tested whether one Agent Skill could guide the adaptation of two independently developed solvers with different native adjoint capabilities.
The agent built DiffSPECFEM2D by reusing the native SPECFEM2D adjoint and built DiffAxiSEM3D by adding an adjoint time step to the AxiSEM3D code. Each framework generated its own synthetic observations for the recovery test.

Before inversion, we checked a directional derivative of the mean soft-DTW divergence over all shots for each framework. At the smoothed starting model, we compared the inner product of the $V_{\mathrm P}$ gradient and a test direction with the corresponding centered finite difference. With a maximum velocity perturbation of $1\,\mathrm{m/s}$, the two scalar derivatives differed by $6.45\times10^{-6}$ for DiffSPECFEM2D and $1.36\times10^{-5}$ for DiffAxiSEM3D, relative to the larger absolute derivative in each test.

In 2D, DiffSPECFEM2D recovered the main dipping layers of Marmousi-II \cite{martin2006marmousi} using the existing adjoint and soft-DTW divergence. After 300 updates with synthetic displacement observations, the full-grid $V_{\mathrm P}$ root-mean-square error was 47.3\% lower than in the starting model (Figure~\ref{fig:marmousi2-models300}a,b,e).

In 3D, DiffAxiSEM3D recovered the overall shape of the salt body and part of its velocity contrast in the reduced SEG/EAGE Salt model \cite{aminzadeh1997salt} (Figure~\ref{fig:axisem3d-models125}).
The same adjoint time step used to verify the gradient with a squared $L_2$ misfit also supported inversion with soft-DTW divergence. After 125 updates, the soft-DTW divergence was 86.1\% lower than at the start.

Using the same Skill, we adapted both solvers for inversion while retaining their native wave propagation and parallel execution. Both packages accept user-defined differentiable misfit functions in PyTorch, as illustrated here for $V_{\mathrm P}$ inversion.

%% file: sections/hybridadfwi_checkpoint_comparison.tex
\begin{table}[tbp]
\centering
\caption{One forward and gradient evaluation over all 40 shots on the NVIDIA GB10. Time is the mean $\pm$ sample SD across paired runs. GPU memory is the peak reserved by PyTorch (GiB, SD shown if nonzero). Down arrows indicate shorter runtime or lower memory use relative to ADFWI. Each gradient is the mean over 40 shots. Gradient difference is the largest paired relative $L_2$ difference between the complete $V_{\mathrm P}$ gradient vectors, with ADFWI as the reference.
We used three paired runs per setting.}
\label{tab:hybridadfwi-benchmark}
\begingroup
\fontsize{9}{13}\selectfont
\setlength{\tabcolsep}{3pt}
\newcommand{\benchSD}[1]{\hspace{.16667em}{\fontsize{8}{10}\selectfont$\pm\,#1$}}
\newlength{\adMemoryWidth}
\newlength{\hybridMemoryWidth}
\newlength{\reductionWidth}
\settowidth{\adMemoryWidth}{\(00.00\)}
\settowidth{\hybridMemoryWidth}{\(\mathbf{00.00}\)}
\newcommand{\benchDecrease}[2]{(\(#1{#2}\,{\downarrow}\))}
\settowidth{\reductionWidth}{\benchDecrease{00.00}{\%}}
\newcommand{\adMemory}[1]{\makebox[\adMemoryWidth][r]{\(#1\)}}
\newcommand{\hybridMemory}[2]{%
  \makebox[\hybridMemoryWidth][r]{\(\mathbf{#1}\)}\enspace
  \makebox[\reductionWidth][r]{\benchDecrease{#2}{\%}}}
\begin{tabular}{@{}c r@{}l c r@{}l c c@{}}
\toprule
\multirow{2}{*}[-0.2em]{\shortstack{Batch\\size}} & \multicolumn{3}{c}{ADFWI} & \multicolumn{3}{c}{HybridADFWI} & \multirow{2}{*}[-0.2em]{\shortstack{Gradient\\difference}} \\
\cmidrule(lr){2-4}\cmidrule(lr){5-7}
& \multicolumn{2}{c}{Time (s)} & GPU memory & \multicolumn{2}{c}{Time (s)} & GPU memory & \\
\midrule
\multicolumn{8}{@{}l@{}}{Checkpointing enabled} \\
40 & $202.78$ & \benchSD{0.06} & \adMemory{6.36} & $\mathbf{63.95}$ & \benchSD{0.03}\enspace\benchDecrease{3.17}{\times} & \hybridMemory{5.89}{7.25} & \(5.27\times10^{-7}\) \\
20 & $195.67$ & \benchSD{0.90} & \adMemory{2.89} & $\mathbf{55.40}$ & \benchSD{0.24}\enspace\benchDecrease{3.53}{\times} & \hybridMemory{2.67}{7.44} & \(7.34\times10^{-7}\) \\
1 & $1015.50$ & \benchSD{0.96} & \adMemory{1.82} & $\mathbf{420.81}$ & \benchSD{0.36}\enspace\benchDecrease{2.41}{\times} & \hybridMemory{1.73}{5.05} & \(4.73\times10^{-7}\) \\
\midrule
\multicolumn{8}{@{}l@{}}{Checkpointing disabled} \\
40 & $185.32$ & \benchSD{0.80} & \adMemory{35.73} & $\mathbf{49.71}$ & \benchSD{0.56}\enspace\benchDecrease{3.73}{\times} & \hybridMemory{29.34}{17.88} & \(4.45\times10^{-7}\) \\
20 & $187.21$ & \benchSD{0.16} & \adMemory{16.76} & $\mathbf{40.62}$ & \benchSD{0.27}\enspace\benchDecrease{4.61}{\times} & \hybridMemory{15.40}{8.08} & \(1.35\times10^{-6}\) \\
1 & $866.82$ & \benchSD{0.29} & \adMemory{2.93} & $\mathbf{286.03}$ & \benchSD{0.16}\enspace\benchDecrease{3.03}{\times} & \hybridMemory{2.28}{22.08} & \(5.00\times10^{-7}\) \\
\bottomrule
\end{tabular}
\endgroup
\end{table}

%% file: sections/discussion.tex
\section{Discussion and conclusions}
\label{sec:discussion}

In this study, we developed a non-intrusive workflow and an Agent Skill for adapting existing high-performance seismic solvers to inversion with user-defined misfit functions.
AD computes only the derivative of the misfit with respect to synthetic data. The solver injects this derivative at the receivers to form the adjoint source and computes the model gradient using its native numerical kernels and parallel execution.
With the same forward solver, HybridADFWI and full AD in ADFWI produced closely matching model gradients and similar inverted velocity models.
Across the tested configurations, HybridADFWI took less time and used less peak reserved GPU memory per forward and gradient evaluation (Table~\ref{tab:hybridadfwi-benchmark}).
We used the same Skill to build DiffSPECFEM2D from SPECFEM2D and DiffAxiSEM3D from AxiSEM3D. Both frameworks passed directional gradient checks. In synthetic inversions, DiffSPECFEM2D recovered the main dipping layers of Marmousi-II, and DiffAxiSEM3D recovered the overall shape of the SEG/EAGE salt body and part of its velocity contrast.

Once a solver has been adapted, researchers can change the waveform misfit and reuse its verified adjoint and sensitivity routines.
For example, switching from squared $L_2$ to soft-DTW divergence in DiffAxiSEM3D changed the adjoint source. The same native routines propagated the new source and computed the model gradient.

For a new solver, the adaptation effort depends on which adjoint and sensitivity routines its code already provides. Given the source code and a working forward example, the agent can reuse routines that match the discrete wave propagation, receiver sampling, and chosen model parameters. It implements any missing operations and checks the resulting model gradient.

The workflow could extend to other inverse problems that already use adjoint models. Examples include estimating ocean circulation \cite{stammer2002global} and inferring surface $\mathrm{CO}_2$ fluxes from satellite measurements \cite{chevallier2005inferring}. Photonic inverse design already combines AD with adjoint simulations \cite{luce2024merging}.
Each application would need a test suited to its scientific goal. In seismology, recovery of known structure provides a direct check that a mature solver can support inversion with a user-defined misfit.

%% file: sections/availability.tex
\section*{Data and Code Availability}
The following GitHub repositories contain the source code for the Agent Skill and the two differentiable inversion frameworks.
\begin{itemize}
  \item Agent Skill for adapting seismic solvers: \url{https://github.com/Wangxi404/differentiate-seismic-solver}
  \item DiffSPECFEM2D: \url{https://github.com/Wangxi404/DiffSPECFEM-native}
  \item DiffAxiSEM3D: \url{https://github.com/Wangxi404/DiffAxiSEM3D-native}
\end{itemize}
Both framework repositories explain how to reproduce the examples and run inversions with user-defined misfit functions.

Each example generates its own synthetic observations from a published reference model.
The 2D example downloads the Marmousi-II $V_{\mathrm P}$, $V_{\mathrm S}$, and density grids \cite{martin2006marmousi} from the \href{https://www.agl.uh.edu/downloads/downloads.htm}{University of Houston AGL Model Definition archive} and verifies their checksums before preparing the inversion inputs.
The 3D example includes a resampled reference model and a smoothed starting model derived from the \href{https://doi.org/10.5281/zenodo.23469308}{SEG/EAGE Salt velocity model} \cite{aminzadeh1997salt}. The Society of Exploration Geophysicists distributes the source model under \href{https://creativecommons.org/licenses/by/4.0/}{CC BY 4.0}.